\documentclass[           
               nofootinbib,         
               tightenlines,        
               floats,floatfix      
               ]{revtex4}

\usepackage{graphicx}
\usepackage{latexsym}
\usepackage{amsmath,amssymb}        
\usepackage[draft=false]{hyperref}

\begin{document}

\title{Gowdy Cosmological Models from Stringy Black Holes}

\author{Tzihu\'e Cisneros--P\'erez}

\author{Alfredo Herrera--Aguilar}

\author{Julio C\'esar Mej\'\i a--Ambriz}

\affiliation{Instituto de F\'{\i}sica y Matem\'{a}ticas, Universidad
Michoacana de San Nicol\'as de Hidalgo \\
Edificio C--3, Ciudad Universitaria, Morelia,
Michoac\'{a}n, CP 58040, M\'{e}xico.\\
E-mails: tzihue@ifm.umich.mx,\hspace{.2cm}
herrera@ifm.umich.mx,\hspace{.2cm} julio@ifm.umich.mx}

\author{Violeta Rojas--Mac\'\i as}

\affiliation{Facultad de Ciencias F\'\i sico--Matem\'aticas,
Universidad Michoacana de San Nicol\'as de Hidalgo\\
Edificio B, Ciudad Universitaria, Morelia, Michoac\'{a}n, CP 58040
M\'exico}

\pacs{98.80.Hw, 04.20.Jb, 04.50.+h, 11.25.Mj}

\begin{abstract}
In the framework of $4D$ Einstein--Maxwell Dilaton--Axion theory we
show how to obtain a family of both unpolarized and polarized
$S^1\times S^2$ Gowdy cosmological models endowed with nontrivial
axion, dilaton and electromagnetic fields from a solitonic rotating
black hole--type solution by interchanging the $r$ and $t$
coordinates in the region located between the horizons of the black
hole configuration. We also get a family of Kantowski--Sachs
cosmologies with topology $R^1\times S^2$ from the polarized Gowdy
cosmological models by decompactifying one of the compact
dimensions.
\end{abstract}

\maketitle

\def\be{\begin{eqnarray}}
\def\ee{\end{eqnarray}}
\def\ba{\begin{array}}
\def\ea{\end{array}}

\section{Introduction}
In the last few years there were many attempts to look at cosmology
from a string theory point of view. Moreover, string cosmology
becomes a subject of great interest among both theoreticians and
phenomenologist. In the present time here are many cosmological
scenarios and the topic itself is experiencing tremendous activity.
For useful reviews see, for instance, \cite{revs} and the references
quoted therein.

In the framework of general relativity, Kantowski and Sachs
\cite{ks} proposed a method which relates the inner region of a
static black hole solutions to an homogeneous cosmological
background under the simple coordinate transformation $r
\longleftrightarrow t$. This result was further generalized and a
relationship between Gowdy cosmologies \cite{gowdy} and the Kerr
rotating black hole was established by Obreg\'on, Quevedo and Ryan
in \cite{oqr}. In this latter case, the coordinate interchange
mentioned above relates the region located between the two
horizons of the rotating black hole solution to a cosmological
model of Gowdy type. Thus, this simple coordinate transformation
enables us to obtain straightforwardly cosmological backgrounds
from black hole configurations, and viceversa, without solving the
Einstein equations, a nontrivial fact which is worth taking into
account. Recently, several papers concerning the physics of
inhomogeneous cosmologies have appeared in the literature in the
framework of Einstein,  Einstein--Maxwell, dilaton gravity, sugra
and string/M theories \cite{inhomcosmol}--\cite{ccq2}.

The idea of this brief report consists of extrapolating this method
to the realm of the $4D$ low--energy heterotic string theory which
describes gravity coupled to a dilaton, an axion and just one
electromagnetic vector field\footnote{A stringy Kantowski--Sachs
cosmological solution describing gravity coupled to dilaton and
Kalb--Ramond (axion) fields was reported in \cite{ksbd}.}. This
generalization becomes possible in the case when the $4D$ theory
possesses two commuting Killing vectors (as it takes place within
the framework of general relativity) and can be lifted to any
dimensions for configurations which posses $D-2$ commuting Killing
vectors.

Thus, in this work we shall perform a straightforward
implementation of the coordinate transformation
$r\longleftrightarrow t$ in order to obtain several families of
cosmological backgrounds from black hole configurations (and
viceversa) without tears. Namely, by starting with a rotating
field configuration of black hole type possessing two horizons, we
apply such a coordinate interchange in the region located between
the horizons, and as a result we get as well several families of
inhomogeneous cosmological model of Gowdy type, both polarized and
unpolarized. It is interesting to point out that a family of
Kantowski--Sachs cosmologies arises from the polarized Gowdy
cosmological models by decompactifying one of the compact
dimensions.

\section{The General Relativity Side of the Story}

We start this Section by quoting the Schwarzschild black hole
solution to the vacuum Einstein's equations \be
ds^2=-(1-2m/r)dt^2+(1-2m/r)^{-1}dr^2+r^2(d\theta^2+\sin^2\theta\
d\varphi^2),\label{sch}\ee where $m$ is a constant parameter which
can be interpreted as the mass of the static black hole.

On the other side, we shall recall as well a particular cosmological
model which has been reported for the first time in \cite{chernov},
but which is quoted in the literature as the Kantowski--Sachs
cosmologies \cite{ks}, namely, a metric of the form \be
ds^2=-N(t)^2dt^2+e^{2\sqrt{3}\beta(t)}dr^2+e^{-2\sqrt{3}\beta(t)}
e^{-2\sqrt{3}\Omega(t)}(d\theta^2+\sin^2\theta\
d\varphi^2).\label{ks}\ee

Kantowski and Sachs obtained the following one parameter family of
solutions to the vacuum Einstein's equations \be N(t)^2 =
\left(\frac{\alpha}{t}-1\right)^{-1}, \qquad e^{2\sqrt{3}\beta(t)}
= \frac{\alpha}{t}-1, \qquad e^{-2\sqrt{3}\Omega(t)} =
t^2\left(\frac{\alpha}{t}-1\right).\label{kssol}\ee As it was
mentioned above, they also realized that this homogeneous
cosmological model is related to the Schwarzschild black hole
solution inside the horizon under the coordinate transformation $r
\longleftrightarrow t$, i.e., the solutions (\ref{sch}) and
(\ref{ks})--(\ref{kssol}) map into each other by interchanging the
coordinates $r$ and $t$ in the region where $r<2m$ with the
following identification $\alpha=2m$.

Taking advantage of this fact, Obreg\'on, Quevedo and Ryan
implemented this coordinate map in the Kerr solution and showed that
the metric of a spinning black hole between the inner and outer
horizons can be reinterpreted as an exact cosmological solution of
Gowdy type with topology $S^1\times S^2$ (for details see
\cite{oqr}).

In order to see how this fact takes place in this case, let us
express the Kerr metric in the Boyer--Lindquist coordinates \be
ds^2=
-f\left(dt+\frac{2m'ra'\sin^2\theta}{\Delta'-a'^2\sin^2\theta}\
d\varphi\right)^2+f^{-1}
\left[\left(\Delta'\!-\!a'^2\sin^2\theta\right)\left(\frac{dr^2}{\Delta'}+
d\theta^2\right)\!+\!\Delta'\sin^2\theta d\varphi^2\right],
\label{kerr}\ee where \be
f(r)=\frac{\Delta'-a'^2\sin^2\theta}{r^2+a'^2\cos^2\theta}, \qquad
\Delta'(r)= (r-m')^2-\sigma'^2,\qquad \sigma'^2=m'^2-a'^2,\ee and
$m'$ and $a'$ are two constants that represent the mass and the
rotation parameter of the rotating black hole.

This solution possesses two horizons given by the following
expressions \be r_{\pm}=m'\pm\sqrt{m'^2-a'^2}, \label{hors}\ee where
$r_+$ and $r_-$ are called outer and inner horizons, respectively.

On the other side, the unpolarized Gowdy cosmological models with
topology $S^1\times S^2$ \cite{gowdy} have the following form \be
ds^2 = e^{(\tau-\lambda)/2}\left(-e^{-2\tau}d\tau^2 +
d\theta^2\right)+ L\sin(e^{-\tau})
e^P\left[\left(d\delta+Qd\varphi\right)^2+e^{-2P}\sin^2\theta
d\varphi^2\right],\label{unpol}\ee where $\lambda$, $P$ and $Q$ are
functions of $\tau$ and $\theta$, and $L$ stands for and arbitrary
constant.

Here we shall tell just a few words concerning the topology of the
metric. It is clear that the $\theta\varphi$ possesses the topology
of the two--sphere $S^2$. Thus, in order to get a metric with the
topology $S^1\times S^2$, the $\delta$ coordinate must be a compact
one. This can be achieved by requiring $0\leq\delta\leq 2\pi$ with
the points $\delta=0$ and $\delta=2\pi$ identified.

In the case when the functions $Q$ is set to zero, one obtains the
so--called polarized cosmological Gowdy models (see as well
\cite{bg}) \be ds^2 = e^{(\tau-\lambda)/2}\left(-e^{-2\tau}d\tau^2 +
d\theta^2\right)+L\sin(e^{-\tau})\left(e^P
d\delta^2+e^{-P}\sin^2\theta d\varphi^2\right).\label{pol}\ee

By performing the coordinate interchange $t \longleftrightarrow r$
in the region located between the inner and outer horizons
$r_-<r<r_+$ of the Kerr solution one gets the following metric \be
ds^2= -f'
\left(dt+\frac{2m'ta'\sin^2\theta}{\Delta'-a'^2\sin^2\theta}\,d\varphi\right)^2+
f'^{-1}
\left[\left(\Delta'\!-\!a'^2\sin^2\theta\right)\left(\frac{dt^2}{\Delta'}+d\theta^2\right)\!+\!
\Delta'\sin^2\theta d\varphi^2\right],\label{gk}\ee where now we
have \be f'(t)=\frac{\Delta'-a'^2\sin^2\theta}{t^2+a'^2\cos^2\theta}
\ee and $$\Delta'(t)= (t-m')^2-\sigma'^2.$$

By performing the following coordinate transformations in this
metric (\ref{gk}) \be
t=\alpha\left[1+\sqrt{1-\beta^2}\cos(e^{-\tau})\right],\quad
r=\delta,\label{risdelta1}\ee where $a'=\alpha\beta$ and
$m'=\alpha,$ we effectively compactify the $r$--coordinate. Thus, we
have a metric with the topology $S^1\times S^2$ of the unpolarized
Gowdy cosmological models. It is an straightforward exercise to
probe that the metric (\ref{gk}) transforms into the metric
(\ref{unpol}) under the following identifications \be
Q=\frac{-2\alpha\beta[1+\sqrt{1-\beta^2}\cos(e^{-\tau})]\sin^2\theta}
{(1-\beta^2)\sin^2(e^{-\tau})+\beta^2\sin^2\theta},\label{Q'}\ee \be
e^P= \frac{(1-\beta^2)\sin^2(e^{-\tau})+\beta^2\sin^2\theta}
{\left\{\left[1+\sqrt{1-\beta^2}\cos(e^{-\tau})\right]^2+\beta^2\cos^2\theta\right\}}
\left[\alpha\sqrt{1-\beta^2}\sin(e^{-\tau})\right]^{-1}, \label{P'}
\ee \be e^{\frac{(\tau-\lambda)}{2}}=
\alpha^2\left\{\left[1\!+\!\sqrt{1\!-\!\beta^2}\cos(e^{-\tau})\right]^2\!
+\!\beta^2\cos^2\theta\right\},\label{lambda'}\ee \be L =
\alpha\sqrt{1-\beta^2}=\sigma'.\label{L'}\ee

In the particular case when the rotation parameter $a'$, in the
framework of the spinning black hole configuration, is set to zero,
the cosmological parameter $\beta$, also disappears and, in turn,
the function $Q$ vanishes as well. As a result one obtains a family
of polarized Gowdy cosmological models with the topology $S^1\times
S^2$.

Here we would like to make an important remark. The vanishing of
the parameter $a'$ or $\beta$, which implies the vanishing of the
function $Q$, leads as well to the vanishing of the inner horizon.
Thus, when performing the coordinate transformation $r
\longleftrightarrow t$, one must take care of mapping the inner
part of the black hole configuration into the polarized Gowdy
cosmological background since we do not have anymore two horizons.

This fact is crucial in order to understand that in this limit, if
we decompactify the $\delta$--coordinate, we recover a metric with
topology $S^2\times R^1$ and, hence, another family of
Kantowski--Sachs cosmological models with the identifications
previously indicated (\ref{kssol}).

\section{The String Theory Side of the Story}

Let us turn our attention to the string theory counterpart of this
story. The $4D$ Einstein-Maxwell theory with dilaton and axion
fields (EMDA) is one of the simplest low--energy string gravity
models. It arises as the corresponding truncation of the critical
heterotic string theory (D=10, with 16 $U(1)$ vector fields)
reduced to four dimensions with no moduli fields excited and just
one non--vanishing vector field. In the Einstein frame it is
described by the action \be S = \int d^4 x |g|^{\frac{1}{2}}\left[
-R + 2(\partial\phi)^2 + \frac{1}{2}e^{4\phi}(\partial \kappa)^2-
e^{-2\phi}F^2 - \kappa F\tilde F\right], \ee where \be
F_{\mu\nu}=\partial_{\mu}A_{\nu}-\partial_{\nu}A_{\mu}, \qquad
\tilde
F^{\mu\nu}=\frac{1}{2}E^{\mu\nu\lambda\sigma}F_{\lambda\sigma},
\nonumber\ee are the strength of the $U(1)$ Maxwell field and its
dual tensor, respectively, $\phi$ is the dilaton field, $\kappa$
is the pseudoscalar axion field, and
$E^{\mu\nu\lambda\sigma}=\epsilon^{\mu\nu\lambda\sigma}/\sqrt{|g|}$.
Formally, the EMDA theory can be considered as an extension of the
Einstein--Maxwell system to the case when one takes into account
the (pseudo)scalar dilaton and axion fields.

It turns out that when this theory admits the presence of two
commuting Killing vectors, the corresponding field equations can be
expressed in a simple chiral form in terms of the so--called matrix
Ernst potentials \cite{mep} and, hence, the inverse scattering
method can be implemented in order to construct exact solutions. By
making use of this method, Yurova has obtained a seven parameter
soliton solution to the field equations of this theory \cite{y}. The
metric of this solution possesses as well the form of a rotating
black hole configuration and is endowed with two horizons.

It is precisely a six parametric subclass of this family of
solutions (we shall set to zero the so--called NUT parameter in
order to restrict ourself to asymptotically flat field
configurations) that we shall use in order to obtain new
cosmological backgrounds. Among these solutions we shall encounter
inhomogeneous unpolarized and polarized Gowdy cosmological models as
well as homogeneous Kantowski--Sachs cosmologies.

Here we won't derive the Yurova's soliton, but we shall just quote
the solution: \be ds^2=-{\cal\widetilde F}
\left[dt+\frac{2a\sin^2\theta[mr-(Q_e^2+Q_m^2)/2]}
{\widetilde\Delta-a^2\sin^2\theta}d\varphi \right]^2 +
{\cal\widetilde F}^{-1}
\left[\left(\widetilde\Delta-a^2\sin^2\theta\right)\left(\frac{dr^2}{\widetilde\Delta}
+d\theta^2\right)+\widetilde\Delta\sin^2\theta d\varphi^2 \right],
\label{metric}\ee where \be {\cal\widetilde
F}(r)=\frac{\widetilde\Delta-a^2\sin^2\theta}{r^2+a^2\cos^2\theta-D^2-K^2},
\ee the function $\widetilde\Delta$ still possesses the same form
\be \widetilde\Delta =(r-m)^2-\tilde\sigma^2,\label{Delta}\ee but
now the constant $\tilde\sigma$ consists of a larger quadratic
combination of constants \be
\tilde\sigma^2=m^2+D^2+K^2-Q_e^2-Q_m^2-a^2 \label{tsigma}\ee where
the constant parameters have the following physical interpretation:
$m$ denotes the mass of the gravitational configuration, $D$ is the
dilaton charge, $K$ is the axion one, $Q_e$ and $Q_m$ label the
electric and magnetic charges, respectively, and $a$ stands for the
angular momentum per unit mass of the rotating soliton.

This solution possesses as well inner and outer horizons defined
through the relations \be r_{\pm}=m\pm\tilde\sigma.\label{rpm}\ee

The expressions for the matter fields in the black hole picture are
the following:

\noindent the dilaton field is \be
e^{2\phi}=\frac{(r+D)^2-(K+a\cos\theta)^2}{r^2+a^2\cos^2\theta-D^2-K^2},\label{phi}\ee
and the axion field reads
\be\kappa=\frac{2(Kr-Da\cos\theta)}{(r+D)^2-(K+a\cos\theta)^2},\label{axion}\ee
whereas the electric and magnetic potentials adopt the form \be
v=\frac{\sqrt2(Q_er+Q_ma\cos\theta+DQ_e+KQ_m)}{r^2+a^2\cos^2\theta-D^2-K^2},\label{ep}\ee
\be
u=\frac{\sqrt2(Q_mr-Q_ea\cos\theta+DQ_m-KQ_e)}{r^2+a^2\cos^2\theta-D^2-K^2},\label{mp}\ee

In the same spirit as in the Kerr solution we perform the coordinate
transformation $t\longleftrightarrow r$ in  the region located
between the two horizons and obtain the following gravitational
field configuration \be ds^2=-{\cal F}
\left[dr+\frac{2a\sin^2\theta[mt-(Q_e^2+Q_m^2)/2]}
{\Delta-a^2\sin^2\theta}d\varphi \right]^2 + {\cal F}^{-1}
\left[\left(\Delta-a^2\sin^2\theta\right)\left(\frac{dt^2}{\Delta} +
d\theta^2\right) + \Delta\sin^2\theta d\varphi^2 \right],
\label{metrict}\ee where now \be{\cal
F}(t)=\frac{\Delta-a^2\sin^2\theta}{t^2+a^2\cos^2\theta-D^2-K^2},
\ee and \be\Delta(t)=(t-m)^2-\tilde\sigma^2.\ee

In order to interpret this metric as an inhomogeneous unpolarized
Gowdy cosmological model with topology $S^1\times S^2$, we again
perform a coordinate transformations of the form (\ref{risdelta1})
\be t=\alpha+\sigma\cos(e^{-\tau}), \qquad r\equiv\delta, \qquad
m=\alpha, \label{risdelta}\ee compactifying in this way the
$r$--coordinate, and make the following identifications for the
functions $Q$, $P$ and $\lambda$ \be Q =
\frac{-2\gamma\sin^2\theta\left\{{\alpha\left[\alpha+\sigma\cos\left(e^{-\tau}\right)\right]
-\left(q_e^2+q_m^2\right)/2}\right\}}
{\sigma^2\sin^2(e^{-\tau})+\gamma^2\sin^2\theta},\label{Q}\ee \be
e^P=\frac{\sigma^2\sin^2\left(e^{-\tau}\right)+\gamma^2\sin^2\theta}
{\left\{\left[\alpha+\sigma\cos\left(e^{-\tau}\right)\right]^2+
\gamma^2\cos^2\theta-d^2-k^2\right\}}
\left[\sigma\sin\left(e^{-\tau}\right)\right]^{-1} \label{P}\ee \be
e^{(\tau-\lambda)/2}=\left[\alpha+\sigma\cos\left(e^{-\tau}\right)\right]^2
+\gamma^2\cos^2\theta-d^2-k^2,\label{lambda}\ee and for the constant
parameter $L$ \be
L=\sigma=\sqrt{\alpha^2+d^2+k^2-q_e^2-q_m^2-\gamma^2},\label{L}\ee
where we have introduced the new charges of the system $D=d$, $K=k$,
$Q_e=q_e$, $Q_m=q_m$, and, finally, the new parameter $a=\gamma$.

The corresponding expressions for the matter fields in the
cosmological framework are:

\noindent the dilaton field reads \be
e^{2\phi}=\frac{\left\{\left[\alpha+\sigma\cos\left(e^{-\tau}\right)\right]
+d\right\}^2-(k+\gamma\cos\theta)^2}
{\left[\alpha+\sigma\cos\left(e^{-\tau}\right)\right]^2+\gamma^2\cos^2\theta-d^2-k^2},
\label{phic}\ee and the axion field is
\be\kappa=\frac{2\left\{k\left[\alpha+\sigma\cos\left(e^{-\tau}\right)\right]
-d\gamma\cos\theta\right\}}{\left\{\left[\alpha+\sigma\cos\left(e^{-\tau}\right)\right]
+d\right\}^2-(k+\gamma\cos\theta)^2},\label{axionc}\ee whereas the
electric and magnetic potentials are given by the following
relations \be
v=\frac{\sqrt2q_e\left[\alpha+\sigma\cos\left(e^{-\tau}\right)\right]}
{\left[\alpha+\sigma\cos\left(e^{-\tau}\right)\right]^2+\gamma^2\cos^2\theta-d^2-k^2}+
\frac{\sqrt2\left(q_m\gamma\cos\theta+dq_e+kq_m\right)}
{\left[\alpha+\sigma\cos\left(e^{-\tau}\right)\right]^2+\gamma^2\cos^2\theta-d^2-k^2},
\label{epc}\ee \be u=
\frac{\sqrt2q_m\left[\alpha+\sigma\cos\left(e^{-\tau}\right)\right]}
{\left[\alpha+\sigma\cos\left(e^{-\tau}\right)\right]^2+\gamma^2\cos^2\theta-d^2-k^2}-
\frac{\sqrt2\left(q_e\gamma\cos\theta-dq_m+kq_e\right)}
{\left[\alpha+\sigma\cos\left(e^{-\tau}\right)\right]^2+\gamma^2\cos^2\theta-d^2-k^2},
\label{mpc}\ee respectively.

As it takes place in the case of the Kerr metric, if we set to zero
the rotation parameter $a$ or, equivalently, the cosmological
parameter $\gamma$, which in turn yields to a vanishing function
$Q$, we are lead to a family of polarized Gowdy cosmological models
with the topology $S^1\times S^2$.

However, it is worth noticing that within the framework of string
theory, the vanishing of the function $Q$ does not yield to the
vanishing of the inner horizon. This fact is due to the presence of
the matter fields, which make their contribution to the size of the
region located between the horizons. Thus, by looking at the
expression of the horizons (\ref{rpm}) we see that the region
located between them gets bigger because of the minus sign of the
$\gamma^2$ term in the definition (\ref{L}) of the constant
$\sigma$.

If we indeed decompactify the $\delta$--coordinate, we obtain a
manifold with topology $R^1\times S^2$ and, hence, a new family of
Kantowski--Sachs cosmological models with the following
identifications for the components of the metric tensor

\be N(t)^2 =\left(\frac{(t-\alpha)^2-\sigma^2}{d^2+k^2-t^2}
\right)^{-1}, \nonumber\ee\be e^{2\sqrt{3}\beta(t)}=
\frac{(t-\alpha)^2-\sigma^2}{d^2+k^2-t^2}, \nonumber\ee\be
e^{-2\sqrt{3}\Omega(t)}=
-\left[(t-\alpha)^2-\sigma^2\right].\label{ksst}\ee On the other
side, the dilaton field is given by \be
e^{2\phi(t)}=\frac{(t+d)^2-k^2}{t^2-d^2-k^2},\label{phiks}\ee and
the axion field reads
\be\kappa(t)=\frac{2kt}{(t+d)^2-k^2},\label{axionks}\ee whereas
the electric and magnetic potentials adopt the form \be
v(t)=\frac{\sqrt2(q_et+dq_e+kq_m)}{t^2-d^2-k^2},\label{epks}\ee
\be
u(t)=\frac{\sqrt2(q_mt+dq_m-kq_e)}{t^2-d^2-k^2},\label{mpks}\ee

A remarkable feature of this latter cosmological Kantowski--Sachs
solution is that all the matter fields decay to zero as far as we
approach the limit $t\longrightarrow\infty$. Thus, their physical
relevance takes place in a interval where $t$ is finite and becomes
crucial as we approach the singularity $t\longrightarrow0$, i.e. in
the early universe.

\section{Discussion}

In this work we have presented an implementation of the coordinate
transformation $t\longleftrightarrow r$ in order to obtain new
(unpolarized and polarized) Gowdy and Kantowski--Sachs cosmological
models in the framework of the $4D$ low--energy effective field
theory of the heterotic string, the so--called EMDA theory. These
stringy cosmological solutions display a different structure with
respect to its general relativity counterparts. In particular, the
behaviour of the horizons of the generated backgrounds turns out to
be very different for the field solutions of the string cosmologies.
This is an interesting subject which deserves further investigation
and will be pursued elsewhere.

There are more directions in which the coordinate exchange presented
in this work can be exploited. For instance, this idea can be easily
generalized to models which involve more than four space--time
dimensions with the aid of the so called matrix Ernst potentials
\cite{mep}, for instance, one could take as starting solution the
field configuration obtained in \cite{iwp}, apply the coordinate
transformation $r\longleftrightarrow t$ in the region located
between the horizons and get multi--dimensional ($D>4$)
inhomogeneous cosmological models of Gowdy type. Another issue
concerns the inclusion of the moduli fields that come from the extra
dimensions. We hope to develop some research along these lines in
the near future.

\section*{Acknowledgments}
One of the authors (AHA) is really grateful to the Theoretical
Physics Department of the Aristotle University of Thessaloniki and,
specially, to Prof. J.E. Paschalis for useful discussions and for
providing a stimulating atmosphere while part of this work was being
done. He also thanks Prof. O. Obreg\'on for fruitful discussions and
Prof. H. Quevedo for an illuminating correspondence on the subject.
Finally, he acknowledges a grant for postdoctoral studies provided
by the Greek Government. This research was supported by grants
CIC-UMSNH-4.18 and CONACYT-42064-F.


\end{document}